\documentclass[a4paper,12pt]{article}
\usepackage[pctex32]{graphics}
\usepackage{amssymb,amsmath}
\usepackage{amsmath,amssymb}
\usepackage{latexsym}
\usepackage{epsfig}
\usepackage[english]{babel}

\newcommand{\be}{\begin{equation}}
\newcommand{\ee}{\end{equation}}
\newcommand{\ba}{\begin{eqnarray}}
\newcommand{\ea}{\end{eqnarray}}

\begin{document}

\begin{titlepage}

\vspace{5mm}
\begin{center}

{\Large \bf van Dam-Veltman-Zakharov discontinuity  in
topologically new massive gravity }

\vskip .6cm

\centerline{\large
 Yun Soo Myung$^{a}$}

\vskip .6cm

{Institute of Basic Science and School of Computer Aided Science,
\\Inje University, Gimhae 621-749, Korea \\}

\end{center}

\begin{center}
\underline{Abstract}
\end{center}

We study van Dam-Veltman-Zakharov discontinuity in the topologically
new massive gravity (TNMG). The reduction from 2 degrees of freedom
to one is interpreted as  van Dam-Veltman-Zakharov discontinuity
appeared when going from  anti-de Sitter spacetime to  Minkowski
spacetime in the linearized TNMG. \vskip .6cm


\vskip 0.8cm

\vspace{15pt} \baselineskip=18pt

\noindent $^a$ysmyung@inje.ac.kr \\

\thispagestyle{empty}
\end{titlepage}

\newpage
\section{Introduction}

Recently, there is a debate on the local degrees of freedom (DOF) of
graviton on  the topologically new massive gravity
(TNMG)~\cite{Andringa:2009yc}.  It turned out that the linearized
TNMG provides  a single spin-2 mode with mass $\frac{m^2}{\mu}$ in
the Minkowski spacetime~\cite{Andringa:2009yc,Dalmazi:2009pm}.
However, it was shown that when using the Hamiltonian formulation
where non-linear effect is not neglected, its DOF is
2~\cite{Hohm:2012vh}. Thus, the reduction (2 $\to$ 1) of local DOF
is considered as an artefact of the linearized approximation in the
Minkowski spacetime.  It implies that the TNMG suffers from a
linearization instability. This reduction is related to  the
emergence of a linearized Weyl (conformal) symmetry at the
linearized TNMG~\cite{Andringa:2009yc}.

On the other hand, there were debates  in the massless limit of the
massive graviton propagator in four dimensional Einstein
gravity~\cite{Hig,KMP,POR,GNie,DW}. An important  issue of this
limit indicates  that van Dam-Veltman-Zakharov (vDVZ)
discontinuity~\cite{DVZ} is peculiar to Minkowski spacetime, but it
seems unlikely to arise in (anti) de Sitter space. The vDVZ
discontinuity implies that the limit of $M^2 \to 0$ does not yield a
massless graviton at the linearized  level. For this purpose,  one
may use  the Fierz-Pauli mass term with mass squared $M_{\rm
FP}^2$~\cite{FP}.

 If the cosmological constant ($\Lambda$) was
introduced~\cite{POR}, a smooth $M_{\rm FP}^2/\Lambda \to0$ limit
exists.
 We note that $M_{\rm
FP}^2 \to0$ and $\Lambda \to 0$ limits do not commute. Taking the
$M_{\rm FP}^2 \to 0$ limit first  and then the $\Lambda \to 0$ limit
recovers a massless graviton with 2 DOF, leading to no vDVZ
discontinuity in the Einstein gravity. Taking the $\Lambda \to 0$
limit first, one encounters the vDVZ discontinuity with 3 DOF. If
one-loop graviton vacuum amplitude is computed for a massive
graviton~\cite{DDLS}, the discontinuity appears again in anti-de
Sitter spacetime. It may imply that the absence of the vDVZ
discontinuity may be considered as an artefact of the linearized
approximation. Also, there was the Boulware-Deser instability which
states that at the non-linearized level,  a ghost appears again in
the massive gravity theory~\cite{BD}.

In this Report, we show that the reduction in the number of local
DOF  is considered as the  van Dam-Veltman-Zakharov discontinuity in
the linearized TNMG  which appears when going from anti-de Sitter to
Minkowski spacetime. Here we will not consider the issues arising
when interactions are included.

We start with  the cosmological  generalized massive gravity (GMG)
action~\cite{Bergshoeff:2009hq,Bergshoeff:2009aq}
 \be\label{gmg}
 S_{\rm cGMG}=\frac{1}{16\pi G}\int d^3x~\sqrt{-g}\left[\sigma R-2\Lambda_0 +\frac{1}{m^2}K+\frac{1}{\mu}L_{\rm CS}\right],
 \ee
where $K$ ($L_{\rm CS}$) is the new massive gravity (NMG) term (the
gravitational Chern-Simons term) given by
 \ba
 K&=&R_{\mu\nu}R^{\mu\nu}-\frac{3}{8}R^2,\\
 L_{\rm CS}&=&\frac{1}{2}\epsilon^{\mu\nu\rho}\Gamma^\alpha_{\mu\beta}
  \left[\partial_\nu\Gamma^\beta_{\alpha\rho}+\frac{2}{3}\Gamma^\beta_{\nu\gamma}\Gamma^\gamma_{\rho\alpha}\right].
 \ea
Here $m$ and  $\mu$ are the two mass parameters, while $\sigma$ is a
dimensionless sign parameter. Also $\Lambda_0$ is the cosmological
constant, and (\ref{gmg}) leads to GMG without it. The pure-$K$ is
the massless new massive gravity (NMG)~\cite{Deser:2009hb} and the
pure-$L_{\rm CS}$ is the conformal Chern-Simons gravity
(CSG)~\cite{Afshar:2011qw}. The topologically massive gravity (TMG)
is a combination of $\sigma R+L_{\rm CS}$~\cite{Deser:1981wh}. In
the limits of $\sigma \to 0$ and $\Lambda_0 \to 0$, one recovers the
TNMG action~\cite{Andringa:2009yc,Ohta:2011rv}. The limit of $\sigma
\to 0$ is called the cosmological TNMG.

Einstein equation of motion takes the form \be \sigma
G_{\mu\nu}+\Lambda_0
g_{\mu\nu}+\frac{1}{2m^2}K_{\mu\nu}+\frac{1}{\mu}C_{\mu\nu}=0, \ee
where
 \ba
 G_{\mu\nu}&=&R_{\mu\nu}-\frac{1}{2}g_{\mu\nu}R, \\
 K_{\mu\nu}&=&-\frac{1}{2} \nabla^2R g_{\mu\nu}-\frac{1}{2}
 \nabla_\mu\nabla_\nu R +2 \nabla^2 R_{\mu\nu} \nonumber  \\
 &+&4R_{\mu\alpha\nu \beta}R^{\alpha \beta} -\frac{3}{2}
 RR_{\mu\nu}-R_{\alpha \beta}R^{\alpha \beta}
 g_{\mu\nu}+\frac{3}{8}R^2 g_{\mu\nu},
 \ea
and the Cotton tensor is given by
 \be
 C_{\mu\nu}=\epsilon_{\mu}^{~\alpha\beta}\nabla_{\alpha}
 \Big(R_{\beta\nu}-\frac{1}{4} g_{\beta\nu} R\Big).
 \ee In this work,
we consider three dimensional  anti de  Sitter (AdS$_3$) spacetimes
\be \label{sle} ds^2_{\rm AdS}=\bar{g}_{\mu\nu}dx^\mu
dx^\nu=\ell^2\Big(-\cosh^2 \rho d\tau^2+d\rho^2+\sinh^2\rho
d\phi^2\Big). \ee In this case, one finds a relation among $m^2$,
$\Lambda_0$, and $\Lambda$ as
 \be
\Lambda_0=\frac{\Lambda^2}{4m^2}+\sigma \Lambda,~~
 \Lambda=-\frac{1}{\ell^2},
 \ee
 where $\ell$ is the AdS$_3$ curvature radius.
Now, we consider the perturbation \be
g_{\mu\nu}=\bar{g}_{\mu\nu}+h_{\mu\nu} \ee around the AdS$_3$
spacetime. Then, the linearized Einstein equation takes the form \be
\label{lieq} \sigma G^{(1)}_{\mu\nu}+\Lambda_0
h_{\mu\nu}+\frac{1}{2m^2}K^{(1)}_{\mu\nu}+\frac{1}{\mu}C^{(1)}_{\mu\nu}=0,
\ee where their linearized tensors were given in
Ref.\cite{Liu:2009pha}.  Let us choose the transverse-traceless (TT)
gauge
 \be \label{TT}
 \bar{\nabla}_{\mu}h^{\mu\nu}=0,~~h^{\mu}_{~\mu}=0.
 \ee
Upon choosing  the TT gauge, the linearized Einstein equation
becomes the fourth-order  differential equation
 \be\label{4theom}
 (\bar{\nabla}^2-2\Lambda)
 \Bigg[\bar{\nabla}^2h_{\mu\nu}+\frac{m^2}{\mu}\epsilon^{~\alpha\beta}_\mu\bar{\nabla}_\alpha h_{\beta\nu}
 +\left(\sigma m^2-\frac{5}{2}\Lambda\right)h_{\mu\nu}\Bigg]=0.
 \ee
 Introducing four  mutually commuting operators
of
 \ba \label{meq1}
 (D^{L/R})^\beta_\mu &=& \delta^\beta_\mu\pm \ell \epsilon_\mu^{~\alpha\beta}\bar{\nabla}_\alpha,\\
 \label{meq2}~~(D^{m_i})^\beta_\mu &=& \delta^\beta_\mu
            +\frac{1}{m_i}\epsilon_\mu^{~\alpha\beta}\bar{\nabla}_\alpha,~~~(i=1,2),
 \ea
the linearized equation of motion (\ref{4theom}) can be written to
be compactly
 \be \label{geq}
 \Big(D^RD^LD^{m_1}D^{m_2}h\Big)_{\mu\nu}=0.
 \ee
Here, two masses  are given by
 \ba
 m_1&=&\frac{m^2}{2\mu}+\sqrt{\frac{m^4}{4\mu^2}-\sigma m^2-\frac{\Lambda}{2}},\nonumber\\
 m_2&=&\frac{m^2}{2\mu}-\sqrt{\frac{m^4}{4\mu^2}-\sigma
 m^2-\frac{\Lambda}{2}},
 \ea
where their combinations are given by \be
m_1+m_2=\frac{m^2}{\mu},~~m_1m_2=\sigma m^2+\frac{\Lambda}{2}. \ee

Firstly, we consider the case of $\Lambda=0$. This corresponds the
GMG, where two massive modes are propagating on the the Minkowski
spacetime. Their masses $m_1$ and $m_2$ all are  positive for
$\sigma=1$ and $4<m^2/\mu^2$  as they take the forms \be
 m_1=\frac{m^2}{2\mu}\Bigg[1+\sqrt{1-\frac{4\mu^2}{m^2}}\Bigg],~~
 m_2=\frac{m^2}{2\mu}\Bigg[1-\sqrt{1-\frac{4\mu^2}{m^2}}\Bigg].
 \ee
 This means that there is no reduction of local DOF when going from
AdS$_3$ to Minkowski spacetime unless the Einstein-Hilbert action is
omitted.

Secondly, let us consider the case without the Einstein-Hilbert
action by choosing $\sigma=0$, leading to  the cosmological TNMG. If
$\Lambda \not=0$, there is no linearized Weyl symmetry. This means
that the linearized equation (\ref{lieq}) does not exhibit an an
additional gauge symmetry acting on $h_{\mu\nu}: h_{\mu\nu} \to
h_{\mu\nu}+2\Omega \bar{g}_{\mu\nu}-2\ell^2
\bar{\nabla}_\mu\bar{\nabla}_\nu \Omega$. Here $\Omega$ is the
linearized Weyl factor.  Therefore, two masses take the forms
 \ba \label{cmass}
 m_1&=&\frac{m^2}{2\mu}\Bigg[1+\sqrt{1+\frac{2\mu^2|\Lambda|}{m^4}}\Bigg],\nonumber\\
 m_2&=&\frac{m^2}{2\mu}\Bigg[1-\sqrt{1+\frac{2\mu^2|\Lambda|}{m^4}}\Bigg]
 \ea
with $|\Lambda|=1/\ell^2$. It is observed from $
(D^{m_i}h)_{\mu\nu}=0$ that for $2|\Lambda| <m^4/\mu^2$, one has two
massive modes with masses \be \label{mmass} m_1 \simeq
\frac{m^2}{\mu},~~m_2 \simeq -\frac{\mu|\Lambda|}{2m^2} \ee in
AdS$_3$ spacetime. Even though $m_2$ seems to be a tachyonic mass in
the Minkowski spacetime, it might not be  a tachyonic mass in the
AdS$_3$ spacetime. For AdS vacua, it is well-known that unitarity
allows scalar fields to have a negative mass squared if   the
Breitenlohner-Freedman (BF) bound is
satisfied~\cite{Breitenlohner:1982bm}. For a massive scalar
propagation in AdS$_3$ spacetime, this bound is given by  \be M^2
\ge m^2_{\rm BF}=-|\Lambda|=-\frac{1}{\ell^2}. \ee It has been
argued that the same bound could apply to spin-2
propagation~\cite{Carlip:2008jk}. However, this bound is not
applicable to the tensor propagation which satisfies the first-order
equation because of its mass squared form.  Observing the
right-tensor gauge  mode satisfying  $(D^Rh)_{\mu\nu}=0$ in
(\ref{geq})~\cite{Li:2008dq}, we might choose
 the mass bound for the tensor propagation as \be M \ge
-\frac{1}{\ell}=-\sqrt{|\Lambda|}. \ee Hence, if the negative mass
$m_2$ satisfies the inequality \be -\frac{1}{\ell} <m_2 <0, \ee the
massive tensor mode with $m_2$ is not a tachyonic mode in the
AdS$_3$ spacetime.

 We are in a position to
discuss the following three limiting cases: \\
 i) Taking  $\Lambda \to 0$ for keeping $m^4/\mu^2$ fixed leads to one DOF with mass
 \be m_1=\frac{m^2}{\mu}, \ee
which corresponds to a single massive graviton propagating on the
Minkowski spacetime~\cite{Andringa:2009yc}. The $m_2=0$ case
corresponds to a massless gauge degree of freedom. \\  ii) Taking
$\Lambda \to 0$ limit first (and then, the $m^4/\mu^2 \to 0$ limit)
recovers the massless NMG with $m_1=0$ and
$m_2=0$~\cite{Deser:2009hb}, leading to no massive DOF. \\ iii)
Taking $m^4/\mu^2 \to 0$ limit first (and then, the $\Lambda \to 0$
limit) recovers the massless NMG with $m_1=0$ and $m_2=0$, leading
to no massive DOF.   Actually, there is no distinction between ii)
and iii).

 For 4D Einstein
gravity with cosmological constant, one has 2 DOF for a massless
graviton,  while one has 5 DOF for a massive graviton. The reduction
of DOF is from 5 to 3 when taking $\Lambda \to 0$ limit first and
then the $M_{\rm FP}^2 \to 0$ limit, leading to the vDVZ
discontinuity. Another reduction ($5\to 2$) is done by taking
$M_{\rm FP}^2 \to 0$ limit first and then the $\Lambda \to 0$ limit,
leading to no vDVZ discontinuity.

At this stage, we have to define the vDVZ discontinuity differently
in three dimensions.  Although a single spin-2 mode of mass $m_1$
becomes the massless mode of massless NMG in the limit of $\mu \to
\infty$ limit~\cite{Deser:2009hb}, we consider the massive mode as
the physically propagating mode in the Minkowski spacetime.  This is
because there is no massless DOF for graviton  as is shown
$D(D-3)/2|_{D=3}=0$ DOF. In general, the massless modes are pure
gauge,  whereas  the massive modes constitute physical degrees of
freedom.  In this sense, it would be say that the massless NMG (3D
Einstein gravity, CSG) has no physical  DOF in the linearized
approximation.  However, TMG has a single massive DOF, while GMG has
two massive DOF. This means that the massless limit of $m^4/\mu^2
\to 0 $ is not necessary to define the vDVZ discontinuity in the
linearized approximation. We do not need to introduce the propagator
approach to define the vDVZ discontinuity in three dimensions.  What
we want to do is to check what happens for masses in the limit of
$\Lambda \to 0$.  The cosmological TNMG has 2 DOF with different
masses (\ref{mmass}). Also, we note that the cosmological term
``$\Lambda_0 h_{\mu\nu}$" in (\ref{lieq}) breaks a linearized Weyl
symmetry which exists at the linearized TNMG. This explains  a
reason why the cosmological TNMG has two massive modes. This two
massive DOF is reduced to one ($m_1=m^2/\mu,~m_2=0$)  when going
from AdS$_3$ spacetime to Minkowski spacetime $(\Lambda \to 0$).
Hence, it is reasonable to insist that this reduction  is called the
vDVZ discontinuity in three dimensions.  That is, the reduction of
$2\to1$ happens only when changing from AdS$_3$ to Minkowski
spactime. In this case, the linearized Weyl symmetry is restored
and, the partial massless is achieved at the linearized level.

Consequently,  the reduction from 2 degrees of freedom to one could
be  interpreted as  van DVZ discontinuity in the linearized TNMG.
This discontinuity appeared when going from anti-de Sitter to
Minkowski spacetime (cosmological TNMG $\to$ TNMG). Partial massless
of $m_1=m^2/\mu$ and $m_2=0$ appeared in the limit of $\Lambda=0$ is
dubbed the vDVZ discontinuity.

\section*{Acknowledgement}
This work was supported supported by the National Research
Foundation of Korea (NRF) grant funded by the Korea government
(MEST) (No.2012-040499).

\end{document}